# Development and Characterization of a Chest CT Atlas

Kaiwen Xu[a*], Riqiang Gao[a], Mirza S. Khan[b, c, d], Shunxing Bao[a], Yucheng Tang[a], Steve A. Deppen[b], Yuankai Huo[a], Kim L. Sandler[b], Pierre P. Massion[b], Mattias P. Heinrich[e], Bennett A. Landman[a, b]

[a] Department of Electrical Engineering and Computer Science, Vanderbilt University, Nashville TN, USA 37235
[b] Vanderbilt University Medical Center, Nashville, TN, USA 37235
[c] Department of Biomedical Informatics, Vanderbilt University, Nashville, TN, 37212
[d] U.S. Department of Veterans Affairs, Nashville, TN, 37212
[e] Institute of Medical Informatics, University of Lübeck, Lübeck, Germany 23562
(*corresponding author: kaiwen.xu@vanderbilt.edu)

**ABSTRACT**

A major goal of lung cancer screening is to identify individuals with particular phenotypes that are associated with high risk of cancer. Identifying relevant phenotypes is complicated by the variation in body position and body composition. In the brain, standardized coordinate systems (e.g., atlases) have enabled separate consideration of local features from gross/global structure. To date, no analogous standard atlas has been presented to enable spatial mapping and harmonization in chest computational tomography (CT). In this paper, we propose a thoracic atlas built upon a large low dose CT (LDCT) database of lung cancer screening program. The study cohort includes 466 male and 387 female subjects with no screening detected malignancy (age 46-79 years, mean 64.9 years). To provide spatial mapping, we optimize a multi-stage inter-subject non-rigid registration pipeline for the entire thoracic space. Briefly, with 50 scans of 50 randomly selected female subjects as fine tuning dataset, we search for the optimal configuration of the non-rigid registration module in a range of adjustable parameters including: registration searching radius, degree of keypoint dispersion, regularization coefficient and similarity patch size, to minimize the registration failure rate approximated by the number of samples with low Dice similarity score (DSC) for lung and body segmentation. We evaluate the optimized pipeline on a separate cohort (100 scans of 50 female and 50 male subjects) relative to two baselines with alternative non-rigid registration module: the same software with default parameters and an alternative software. We achieve a significant improvement in terms of registration success rate based on manual QA. For the entire study cohort, the optimized pipeline achieves a registration success rate of 91.7%. The application validity of the developed atlas is evaluated in terms of discriminative capability for different anatomic phenotypes, including body mass index (BMI), chronic obstructive pulmonary disease (COPD), and coronary artery calcification (CAC).

**Keywords:** Chest CT, Thorax Atlas, Standardized Coordinate System, Non-rigid Registration, Missing Data

## 1. INTRODUCTION

Standardized coordinate systems (e.g., atlases) are a widely used spatial normalization technique in the field of medical image analysis for anatomic variation analysis and outlier detection [1][2]. After projecting samples into a common space, corresponding anatomic locations can be addressed uniformly across the study population. This enables both local and global comparison of tissue intensity and morphology variations. While atlas techniques are widely studied for the brain, only a few research works have been deployed in relation to atlas building in thoracic space. [3] discussed the construction of a normal lung atlas with a small computational tomography (CT) dataset of 20 normal subjects. [4] discussed a normalized thoracic affine coordinate system based on linear morphology geometric features, e.g. planes and directions defined by key points of thoracic cage. None of these works have addressed the task to build atlases for the entire thoracic space using a large screening CT database.

A registration pipeline that can find spatial anatomic correspondence between scans is the core component of creating such a standardized coordinate system. Compared to the brain, the registration in thoracic space is more challenging due to large anatomic variation and deformation. The problem is further complicated by the sliding effect and contrast change between tissues introduced by respiratory motions. In recent years, intensity-driven discrete optimization based registration strategies [5][6][7] have shown promising results in the registration of lungs and the registration of abdominal space [8]. Compared to the traditional continuous optimization approaches, these new strategies are more likely to avoid the local minimum problem and capture the optimal correspondence with large spatial deviations. Moreover, unlike in the brain or lung registration where the field of view (FOV) of the scans can always cover the anatomic region of interest (ROI) (brain or lung), in registration of the entire thoracic space the FOV of moving scans may only provide partial information in correspondence to the anatomic ROI defined by the FOV of the reference scan. The situation is common in screening CT scans due to both positioning and anatomic variations. This leads to the problem of registration with partial or missing information. As shown in [9][10], an iterative multi-stage procedure is a typical method to solve the problem, and a dynamically updating effective registration mask is needed to limit the registration only into the region where valid information can be provided on both moving and reference scans. Among the public released registration tools, *corrField* [6][7] provides both the capability to specify a mask to guide the registration and the effective discrete optimization based registration approach that is suitable for thoracic space.

In this work, we propose an optimized registration pipeline for finding spatial correspondence of the entire thoracic space (Figure 1). The pipeline consists of a preprocessing module, an affine registration module based on *NiftyReg* [11] followed by a non-rigid registration module developed with *corrField*. Compared to the affine module, the non-rigid module is more critical and challenging in providing accurate localized correspondence. In the original paper [6], *corrField* is designed for intra-subject registration in estimation of lung motions for patients with chronic obstructive pulmonary disease (COPD). A registration mask for the lung region is specified to guide the generation of key points for driving the registration, which achieved significant improvement in terms of alignment of lung tissues. However, the method cannot be applied directly to our work. We optimize *corrField* on the following three aspects to fit the task for inter-subject registration of entire thoracic space: (1) introduction of a multistage coarse-to-fine approach to cover large inter-subject deformation and provide an iterative procedure to solve the ROI mismatch problem; (2) replacement of the registration mask of lung region with a dynamically updating effective registration region mask considering the deformed irregular FOV of moving scan; and (3) systematic search of the optimal configurations for the top two most contributing stages to optimize the registration result on a dataset of 50 randomly selected female subjects. In Section 3, we verify the registration accuracy of the pipeline and demonstrates the application validity of the common coordinate space. The overall design of the study is summarized in Fig 1.

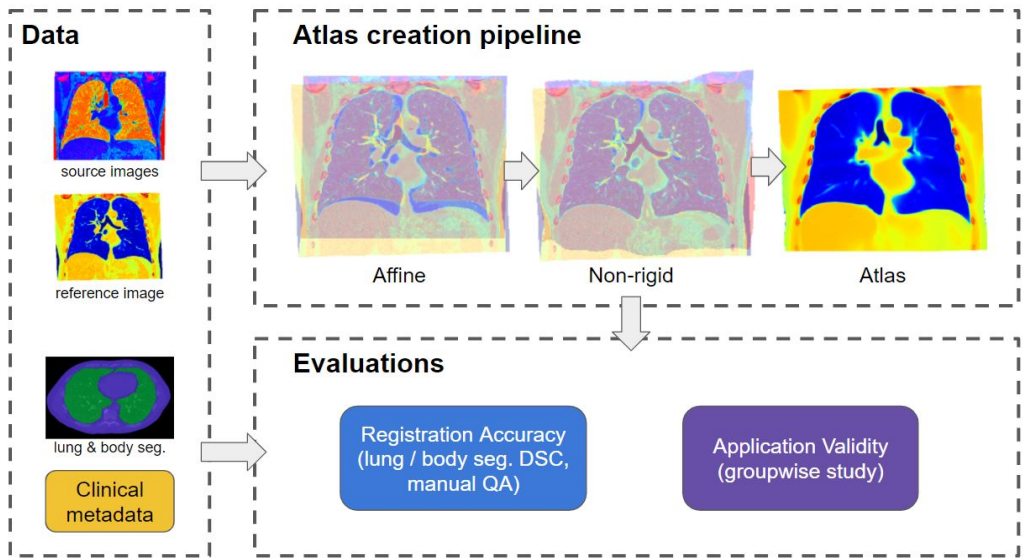

**Figure 1.** Registration pipeline and evaluation methods. Given each scan sample and the reference scan, an affine registration is applied followed by a multi-stage non-rigid registration. Evaluation methods are designed to assess registration accuracy, application validity of the standardized coordinate space.

## 2. METHOD

The atlas creation pipeline consists of four consecutive modules as shown in Fig 2. In the preprocessing step, all image data are first converted into Hounsfield Unit (HU), which is the standard linear density scale for CT images. Then, lung and body segmentations are created based on intensity clipping and morphology operations. To remove noisy ambient information, e.g., the scan table, voxels outside the body segmentation mask are imputed with value -1000 as the HU of air in CT image. The automatically generated lung and body segmentation masks are used in the evaluation of the registration pipeline. (A.2) of Fig 2 shows the overlap of a moving scan with its lung and body segmentation masks. The rest of this section will go through the technical details of affine and non-rigid registration modules and the atlas creation method.

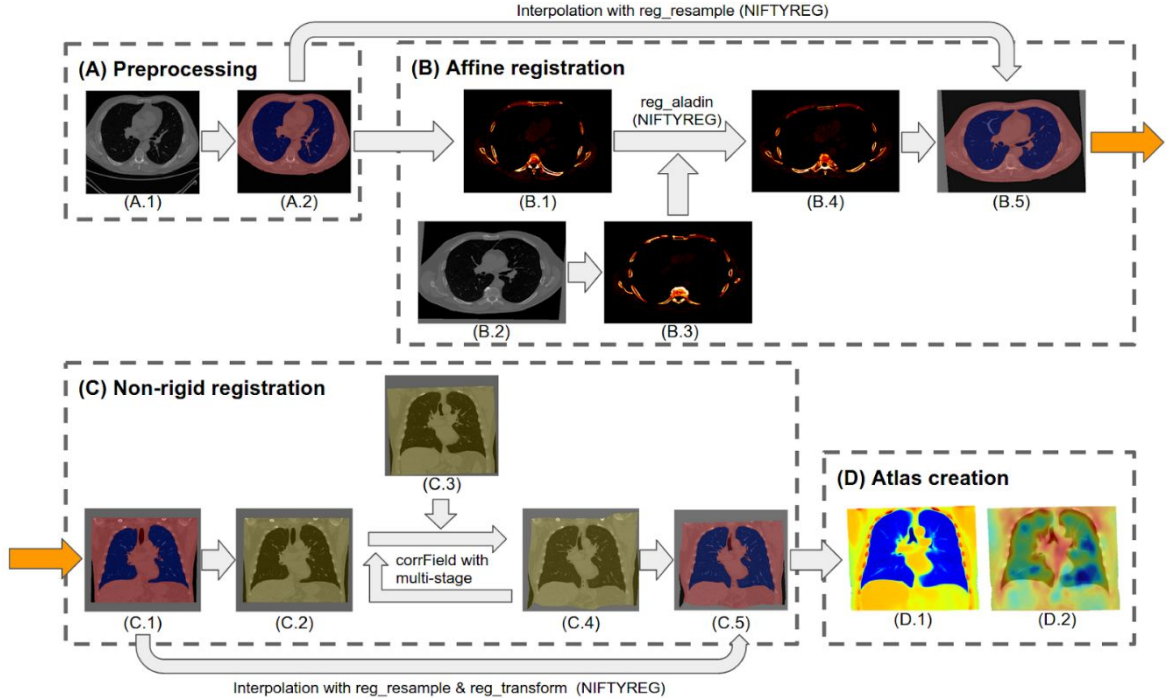

**Figure 2.** Four consecutive modules of the atlas creation pipeline. (A) Preprocessing module: (A.1) original moving scan; (A.2) the overlap between the ambient removed scan and lung & body segmentation masks. (B) Affine registration module based on *NiftyReg* toolbox: (B.1) moving scan with intensity window (0, 1000); (B.2) axial view of reference scan; (B.3) reference scan with intensity window (0, 1000); (B.4) affine registered intensity windowed moving scan; (B.5) interpolation of the preprocessed moving scan with the affine registration transformation matrix. (C) Non-rigid registration module based on *corrField* (registration) and *NiftyReg* (interpolation): (C.1) affine registered moving scan overlapped with lung and body segmentation masks; (C.2) overlapped with non-NaN region mask; (C.3) reference scan with non-NaN region mask; (C.4) non-rigid deformed moving scan overlapped with deformed non-NaN region mask; (C.5) non-rigid deformed moving scan overlapped with deformed lung / body masks. (D) Atlas creation module: (D.1) average atlas of HU; (D.2) average atlas of Jacobian determinant map.

## 2.1 Multi-stage registration pipeline

### 2.1.1 Affine module

[8] shows that the efficiency of non-rigid registration could be significantly impacted by effectiveness of the baseline affine registration. To reduce the large anatomic variation caused by fat tissue, the incoming moving scan is clipped by an intensity window (0, 1000) as in Fig 2 (B.1). The same intensity window is applied to the reference scan Fig 2 (B.2). Then, we register the intensity clipped moving scan to the intensity clipped reference scan using the affine registration module (*reg_aladin*) of the *NiftyReg* toolbox. Fig 2 (B.4) shows the result of registration. Using *reg_resample* of *NiftyReg*, the generated affine transformation is further propagated to the intensity map, lung mask, and body mask. An illustration of the propagated result is shown in Fig 2 (B.5).

### 2.1.2 Non-rigid module

After global alignment with affine registration, large local anatomic differences still exist between the warped scans. To further enable accurate localized correspondence, we designed a multi-stage non-rigid registration module by extending the intrinsic two-stage configuration of *corrField*. The pipeline follows a coarse-to-fine procedure, where a large scale search with strict regularization penalty is first employed followed by localized high-accurate modifications with smaller search range. The final optimized pipeline consists of four major stages with two sub-stages for each major stage. The optimized configuration is detailed in Table 1. To achieve better computing efficiency without much loss of accuracy, the first three major stages operate on downsampled scans (2x2x2mm), and the last stage is on high resolution (1x1x1mm). The transformation fields of each stage are concatenated with *reg_transform* of the *NiftyReg* toolbox to form the final correspondence map.

As discussed in the introduction section, the registration is complicated by the irregular field of view of the moving scan. The situation manifests as the existence of voxels with NaN value after affine registration. Non-nan region masks of both moving and reference scans (as in Fig 2. C.2, Fig 2. C.3) are introduced to label the region with valid HU values. In practice, the NaN voxels are imputed with -1000, for *corrField* requires non-NaN voxels as input. This approximation can introduce artificial boundaries between tissue and ambient, which leads to wrong correspondence in boundary regions. To reduce this effect, an effective registration mask is introduced and fed forward to *corrField* to guide the generation of key points to limit the registration into the region with valid intensity.

2.1.3 Search for optimal configuration

The first two stages of the four-stage pipeline are predominant in terms of registration results and are optimized systematically in our study. We consider the following four adjustable parameters: (1) search radius for each key point; (2) sparsity of the distribution of key points; (3) patch size for self-similarity descriptors (SSC) [6]; and (4) degree of regularization penalty. The registration result is evaluated with the registration failure rate which is approximated by the number of samples with low Dice similarity score (DSC) for lung and body segmentation. In this work, we set the thresholds as 0.92 and 0.975 for lung and body, respectively. We apply the grid searching method for hyper parameter tuning considering the highly non-linear nature of the problem. This leads to 375 testing configurations for each stage. Considering the large computational requirement, only 50 randomly selected female subjects from the study dataset are included in this fine-tune process. The fine-tuned configuration is shown in Table 1.

**Table 1.** The optimized configuration of the four-stage pipeline, with two sub-stages for each stage. Resolution: the resolution of scans in millimeters. Search radius: search radius for each key point, in voxel length (spatial length in millimeters). Keypoint dispersion: the average size of the space dominated by each key point with larger size indicating sparser distribution, in voxel length (spatial length in millimeters). Patch radius: patch size in voxel length for self-similarity descriptors, larger size indicating better stability but less accuracy. Regularization: regularization coefficient, larger value for higher regularization penalty.

| Stage | Resolution (mm) | Search radius | Keypoint dispersion | Patch radius | Regularization |
|---|---|---|---|---|---|
| Stage1 | 2×2×2 | 60×30 (120×60mm) | 8×4 (16×8mm) | 6×4 | 1 |
| Stage2 | 2×2×2 | 32×16 (64×32mm) | 7×3 (14×6mm) | 6×4 | 0.7 |
| Stage3 | 2×2×2 | 10×6 (20×12mm) | 6×3 (12×6mm) | 3×2 | 0.5 |
| Stage4 | 1×1×1 | 20×10 (20×10mm) | 10×5 (10×5mm) | 6×4 | 0.1 |

**2.2 Atlas creation**

2.2.1 Selection of reference scan

To create the atlas, a reference scan is needed to specify the normal anatomy of the thorax space. In our study, this reference is selected from the study cohort as a scan without ill-positioning or screening detected pathology. Then, all scans from the study population are registered to this reference with the optimized registration pipeline. Considering the intrinsic anatomic differences between male and female, we separately build the atlas for genders with references from each gender group.

2.2.2 Residual map and log Jacobian determinant map

For each source scan, the deformation fields from the affine and non-rigid modules are concatenated to form a complete correspondence mapping from original space to the common coordinate space. Based on this correspondence mapping, the following two result maps are generated for downstream studies: (1) residual map, where the voxel value represent the HU of the corresponding location of the source scan; (2) log Jacobian determinant map, where the voxel value represents the ratio of volume change of the corresponding location of the source scan. This enables the study of both tissue density and morphology (shape) variations of the study cohort under the common coordinate space. These maps are further trimmed by the body mask of the reference scan. Furthermore, (Fig 2, D.1) and (Fig 2, D.2) show the atlases for residual maps and log Jacobian determinant maps, respectively.

2.2.3  Atlas creation with missing information

The standard ROI for the common coordinate space is defined as the non-NaN region of the reference scan. For each moving scan, after projected to the common coordinate space, the FOV of the scan is deformed and has mismatched regions in respect to the standard ROI. The effective region of this moving scan is defined as the intersection of its deformed FOV mask and the standard ROI mask. For both HU and log Jacobian determinant maps, only information inside the effective region is used for atlas creation. Thus, the value of the voxel $p$ on the atlas $I_{atlas}$ is defined as,

$$I_{atlas}(p) = \frac{1}{|N_p|} \sum_{S \in N_p} I_s(p), \quad (1)$$

where $N_p$ is the subset of scans that have $p$ inside the effective region, and $I_s$ is the map of scan $s$ that correspond to the atlas to create (HU or log Jacobian determinant).

## 3. EXPERIMENTS AND RESULTS

Our study cohort is selected from Vanderbilt Lung Screening Program (VLSP) (https://www.vumc.org/radiology/lung) [12][13][14][15]. The program is designed for annual screening for current or former smokers with high risk for lung cancer. To create a normal space of this population, patients with screening detected malignancy are excluded from the study. The study cohort consists of 1422 chest LDCT scans from 853 de-identified subjects (age 46-79 years, mean 64.9 years). Clinical metadata including body mass index (BMI), diagnosis information for COPD, and severity of coronary artery calcification (CAC) is also included in the evaluation. All data are retrieved under internal review board supervision. We separately build the atlases for male and female populations with the same optimized pipeline. Due to the space limitations, only results for male population are illustrated in Section 3.2.

**Table 2.** Comparison of averaged Dice similarity score (DSC) and success rate on testing dataset (50 male and 50 female subjects)

|  | Male | | | Female | | |
| --- | --- | --- | --- | --- | --- | --- |
|  | Lung DSC | Body DSC | Success % | Lung DSC | Body DSC | Success % |
| DEEDS | 0.9550±0.0140 | 0.9810±0.0088 | 88% | 0.9526±0.0193 | 0.9697±0.0124 | 72% |
| corrField (default) | 0.9379±0.0125 | 0.9907±0.0042 | 88% | 0.9393±0.0145 | 0.9891±0.0057 | 90% |
| corrField (optimized) | 0.9489±0.0169 | 0.9932±0.0037 | 94% | 0.9474±0.0226 | 0.9901±0.0047 | 94% |

**Table 3.** QA result of registration output of entire study cohort

|  | Succeed | Failed | Total | Success % |
| --- | --- | --- | --- | --- |
| Female | 581 | 67 | 648 | 89.7% |
| Male | 764 | 54 | 818 | 93.4% |
| All | 1345 | 121 | 1466 | 91.7% |

### 3.1 Evaluation of registration accuracy

The optimized registration pipeline is evaluated relative to two baseline pipelines with alternative non-rigid registration modules: (1) *DEEDS* [5], which uses similar discrete optimization based registration approach but cannot specify registration mask; (2) single stage *corrField* with the default configuration. The evaluation dataset is a separate dataset from the 50 female hyper-parameter tuning dataset, which consists of 50 randomly selected males and 50 randomly selected females. The results of this evaluation are shown in Table 2, which indicates a significant improvement of the optimized pipeline in comparison to two baselines in terms of registration success rate. The pipeline achieves an acceptable registration success rate of 91.7% for the entire study cohort. Table 3 shows the breakdown of the success rate for male and female. Failed cases are excluded in the remaining part of this evaluation section.

### 3.2 Groupwise comparison in common coordinate space

To assess the application validity of the common coordinate space, atlases of different phenotypic subgroups were created and compared. In this section, we visually evaluate the discriminative capability for the following three pairs of subgroups: (1) BMI subgroups: subjects with BMI in the normal range (18.5 to 24.9 kg/m$^2$) and those defined as obese (BMI > 30 kg/m$^2$); (2) COPD subgroups: subjects with a formal diagnosis of COPD and those without a COPD diagnosis; and (3) CAC subgroups: subjects without CAC or with mild levels of CAC and those with moderate or severe CAC levels. For each of the comparison pairs, we calculate the average / variance maps of intensity (HU) and the average / variance maps of log Jacobian determinant for each subgroup. The difference maps are generated for quantitative evaluation of the differences between subgroups. We hypothesize that, in groupwise comparison, relative voxel-wise value shifting between the subgroups is more informative than the absolute value. For the averaged intensity map, a higher value indicates higher tissue density in the corresponding anatomic location of the subgroup, and vice versa. On the other hand, for the averaged log Jacobian determinant map, a higher value stands for more volume compression during registration, i.e., larger volume in original space, for the corresponding anatomic location of the subgroup, and vice versa.

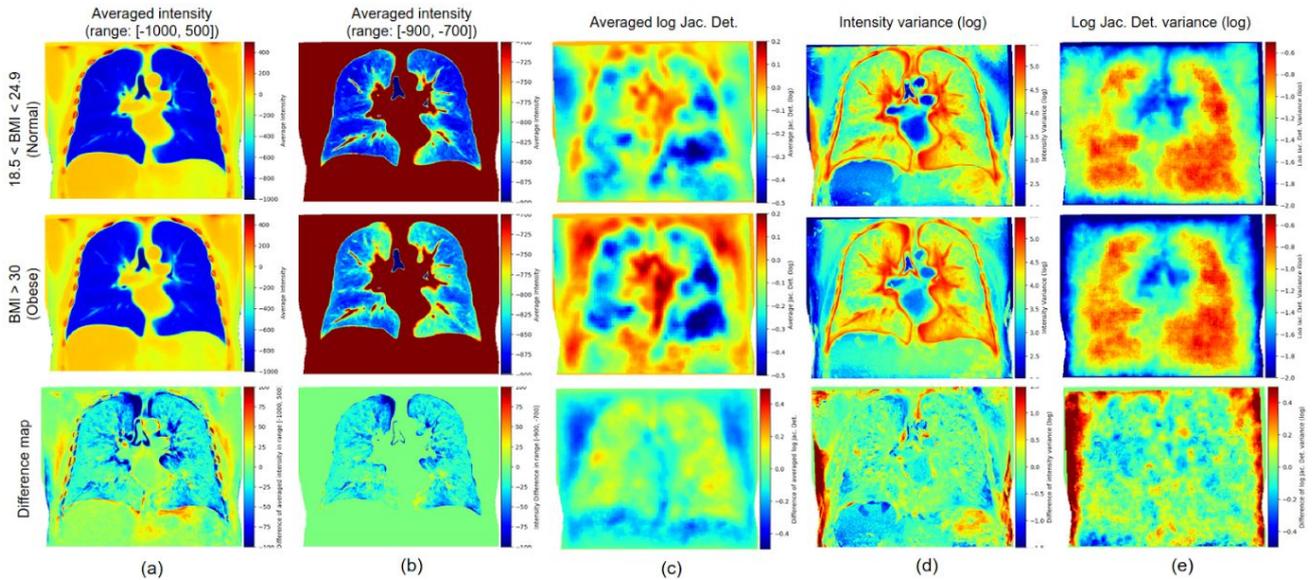

**Figure 3.** Groupwise comparison for normal and obese population. For each column: (a) map of intensity average; (b) map of intensity average clipped with window (-900, -700) HU; (c) average of log Jacobian determinant; (d) variance (log) of intensity; (e) variance (log) of log Jacobian determinant. Top row: population with BMI from 18.5 to 24.9 kg/m$^2$ (normal). Middle row: population with BMI larger than 30 kg/m$^2$ (obese). Bottom row: difference between normal and obese groups.

The results illustrated in Figures 3, 4 and 5 demonstrate the potential application of our thoracic common coordinate space in discriminating the anatomic variation of different clinical conditions. Fig 3 shows the results for BMI subgroups. In column (a) of Fig 3, decreased intensity in the intersection regions of muscle and fat tissues can be observed for the obesity group. In the difference map of the averaged log Jacobian determinant map shown in column (c), we observe increased volume for the thoracic body wall among obese subjects. Fig 4 illustrates the results for COPD subgroups. Decreased intensity of the lung region can be observed for the COPD diagnosed group in column (b), which is the average intensity map shown with a window from -900 to -700 HU. This phenomenon may correspond to the air trapping, small airway disease, and/or lung parenchyma destruction [16][17]. Furthermore, increased lung volume can also be observed in column (c) for COPD patients which may reflect the lung hyperinflation commonly seen in this disease [18]. The results for CAC subgroups are illustrated in Fig 5. Two regions of interest are highlighted: (a) arch of aorta; (b) left coronary artery. Calcium deposits in these two regions are highly correlated and are used as a clinical indicator for cardiovascular diseases [19][20]. As illustrated in the results of both regions, the common coordinate space can capture the difference of the level of calcium deposits between the subgroups.

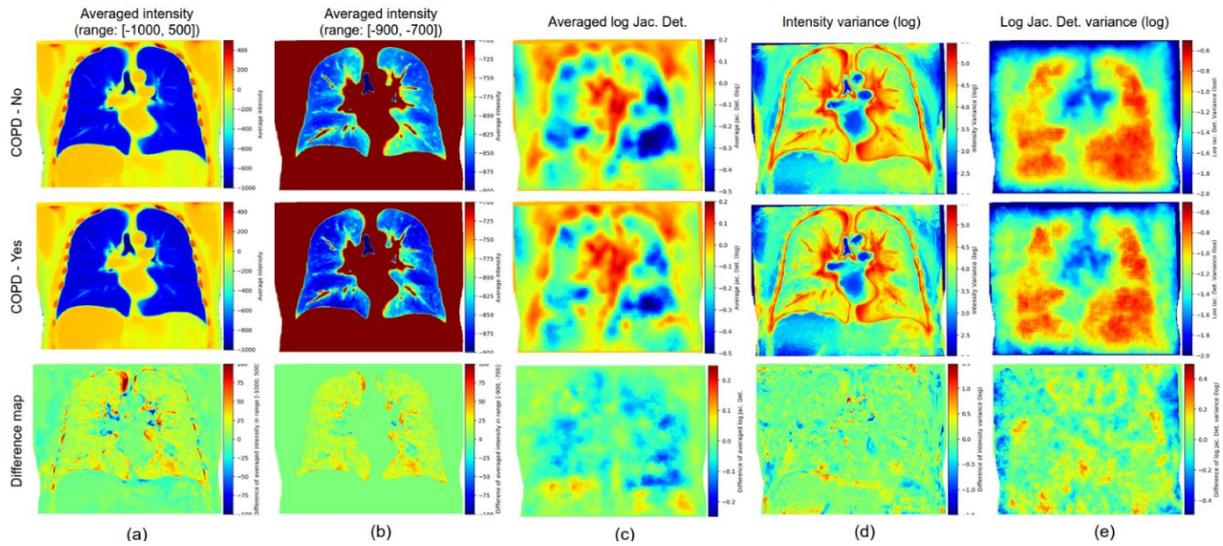

**Figure 4.** Groupwise comparison for population with or without COPD diagnosis. Columns (a)-(e) are with the same definitions as in Fig 2. Top row: population without COPD. Middle row: population diagnosed with COPD. Bottom row: difference between groups with and without COPD.

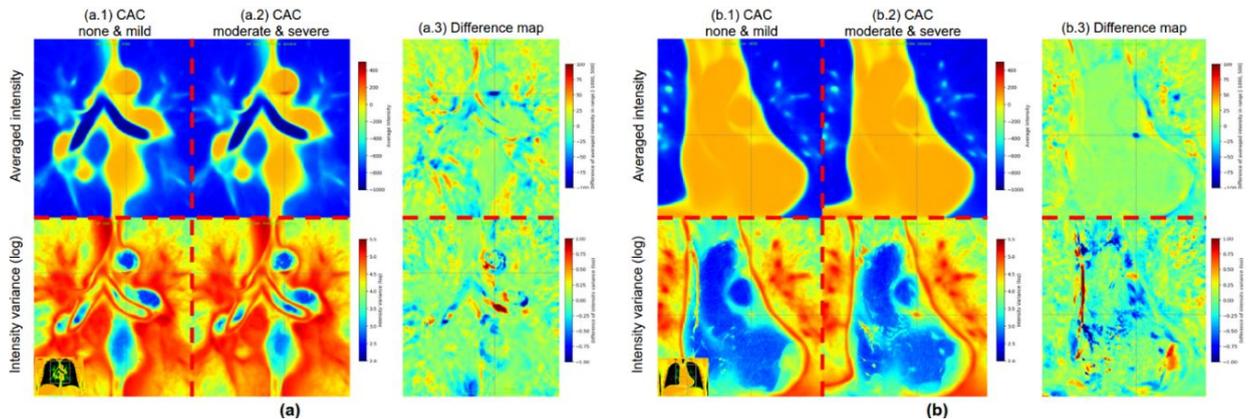

**Figure 5.** Groupwise comparison for population with different severity of coronary artery calcification (CAC). The population is divided into two groups based on the severity of CAC: (1) subjects without CAC or with mild CAC; (2) subjects with moderate or severe CAC. Two ROI regions with high accumulation of calcium are selected for visualization: (a) arch of aorta; (b) left coronary artery. For each group and each ROI region, intensity average maps, and intensity variance maps are shown in columns (a.1, a.2, b.1, b.2). Difference maps of each comparison group are shown in column (a.3) and (b.3).

## 4. CONCLUSION

In this paper, a standardized coordinate system (e.g., atlas) for the entire thoracic space has been built upon a large database of low dose lung screening CT. To achieve this, a multi-stage non-rigid registration pipeline is optimized to find inter-subject correspondence mapping in the entire thoracic space. The pipeline is built based on public available open source registration tools, but with several improvements to achieve acceptable registration accuracy for the new scenario. The optimized pipeline is evaluated relative to two baseline methods. The results show significant improvement in terms of registration success rate. The application validity of the created common coordinate system has been evaluated by groupwise comparison. As demonstrated in the experiments, the groupwise differences for each subgroup pair consist with clinical experience. This indicates the effectiveness of the system in terms of discriminative capability for different anatomic phenotypes.

## ACKNOWLEDGEMENT

This research was supported by NSF CAREER 1452485 and R01 EB017230. This study was supported in part by a UO1 CA196405 to Massion. This study was in part using the resources of the Advanced Computing Center for Research and Education (ACCRE) at Vanderbilt University, Nashville, TN. This project was supported in part by the National Center for Research Resources, Grant UL1 RR024975-01, and is now at the National Center for Advancing Translational Sciences, Grant 2 UL1 TR000445-06. The de-identified imaging dataset(s) used for the analysis described were obtained from ImageVU, a research resource supported by the VICTR CTSA award (ULTR000445 from NCATS/NIH), Vanderbilt University Medical Center institutional funding and Patient-Centered Outcomes Research Institute (PCORI; contract CDRN-1306-04869). This study was funded in part by the Martineau Innovation Fund Grant through the Vanderbilt-Ingram Cancer Center Thoracic Working Group and NCI Early Detection Research Network 2U01CA152662-06.

## REFERENCES

[1] A. Guimond, J. Meunier, and J. P. Thirion, "Average brain models: A convergence study," *Comput. Vis. Image Underst.*, vol. 77, no. 2, pp. 192–210, 2000.

[2] J. E. Rood *et al.*, "Toward a Common Coordinate Framework for the Human Body.," *Cell*, vol. 179, no. 7, pp. 1455–1467, 2019.

[3] B. Li, G. E. Christensen, E. A. Hoffman, G. McLennan, and J. M. Reinhardt, "Establishing a Normative Atlas of the Human Lung," *Acad. Radiol.*, vol. 19, no. 11, pp. 1368–1381, Nov. 2012.

[4] H. Shen and M. Shao, "A thoracic cage coordinate system for recording pathologies in lung CT volume data," *IEEE Nucl. Sci. Symp. Conf. Rec.*, vol. 5, pp. 3029–3031, 2003.

[5] M. P. Heinrich, M. Jenkinson, M. Brady, and J. A. Schnabel, "MRF-Based deformable registration and ventilation estimation of lung CT," *IEEE Trans. Med. Imaging*, vol. 32, no. 7, pp. 1239–1248, 2013.

[6] M. P. Heinrich, H. Handels, and I. J. A. Simpson, "Estimating large lung motion in COPD patients by symmetric regularised correspondence fields," *Lect. Notes Comput. Sci. (including Subser. Lect. Notes Artif. Intell. Lect. Notes Bioinformatics)*, vol. 9350, pp. 338–345, 2015.

[7] J. Ruhaak *et al.*, "Estimation of Large Motion in Lung CT by Integrating Regularized Keypoint Correspondences into Dense Deformable Registration," *IEEE Trans. Med. Imaging*, vol. 36, no. 8, pp. 1746–1757, 2017.

[8] Z. Xu *et al.*, "Evaluation of Six Registration Methods for the Human Abdomen on Clinically Acquired CT," *IEEE Trans. Biomed. Eng.*, vol. 63, no. 8, pp. 1563–1572, Aug. 2016.

[9] S. Periaswamy and H. Farid, "Medical image registration with partial data," *Med. Image Anal.*, vol. 10, no. 3 SPEC. ISS., pp. 452–464, 2006.

[10] N. Chitphakdithai and J. S. Duncan, "Non-rigid Registration with Missing Correspondences in Preoperative and Postresection Brain Images," in *Bone*, vol. 6361, no. 1, T. Jiang, N. Navab, J. P. W. Pluim, and M. A. Viergever, Eds. Berlin, Heidelberg: Springer Berlin Heidelberg, 2010, pp. 367–374.

[11] M. Modat *et al.*, "Fast free-form deformation using graphics processing units," *Comput. Methods Programs Biomed.*, vol. 98, no. 3, pp. 278–284, Jun. 2010.

[12] J. Wang *et al.*, "Lung cancer detection using co-learning from chest CT images and clinical demographics," no. Figure 1, p. 122, 2019.

[13] R. Gao *et al.*, "Deep multi-task prediction of lung cancer and cancer-free progression from censored heterogenous clinical imaging," p. 12, 2020.

[14] R. Gao *et al.*, "Distanced LSTM: Time-Distanced Gates in Long Short-Term Memory Models for Lung Cancer Detection," pp. 1–8.

[15] R. Gao *et al.*, "Time-Distanced Gates in Long Short-Term Memory Networks," *Med. Image Anal.*, p. 101785, Jul. 2020.
[16] Global Initiative for Chronic Obstructive Lung Disease, "Global Strategy for the Diagnosis, Management, and Prevention of Chronic Obstructive Lung Disease: 2019 Report," 2019.
[17] J. Ley-Zaporozhan and E. J. R. Van Beek, "Imaging phenotypes of chronic obstructive pulmonary disease," *J. Magn. Reson. Imaging*, vol. 32, no. 6, pp. 1340–1352, 2010.
[18] G. T. Ferguson, "Why Does the Lung Hyperinflate?," *Proc. Am. Thorac. Soc.*, vol. 3, no. 2, pp. 176–179, Apr. 2006.
[19] J. F. Breen *et al.*, "Coronary artery calcification detected with ultrafast CT as an indication of coronary artery disease.," *Radiology*, vol. 185, no. 2, pp. 435–439, Nov. 1992.
[20] C. Iribarren, S. Sidney, B. Sternfeld, and W. S. Browner, "Calcification of the Aortic Arch," *JAMA*, vol. 283, no. 21, p. 2810, Jun. 2000.